\documentclass[pof,a4paper,superscriptaddress,citesort,twocolumn,10pt]{revtex4}

\usepackage{epsfig}
\usepackage{bm}
\usepackage{graphicx}
 \usepackage{longtable}
\usepackage{latexsym}
\usepackage{color}

\newcommand{\bff}{{\bf f}}
\newcommand{\bfk}{{\bf k}}
\newcommand{\bfx}{{\bf x}}

\newcommand{\bfu}{{\bf u}}
\newcommand{\bfU}{{\bf U}}

\newcommand{\bfr}{{\bf r}}
\newcommand{\bfR}{{\bf R}}
\newcommand{\bfL}{{\bf L}}

\newcommand{\cL}{{L}}
\newcommand{\cE}{{E}}

\begin{document}
% \title{Dynamic Scaling and Multiscaling in Hydrodynamic Turbulence}
\title{Multi-Time Multi-Scale Correlation Functions in Hydrodynamic
  Turbulence}

\author{Luca Biferale}  
% \affiliation{International Collaboration for Turbulence Research}
\affiliation{International Collaboration for Turbulence Research,
  Dept. Physics and INFN, University of Rome ``Tor Vergata'' Via della Ricerca
  Scientifica 1, 00133 Rome, Italy.}

\author{Enrico Calzavarini}  
% \affiliation{International Collaboration for Turbulence Research}
\affiliation{International Collaboration for Turbulence Research,
  Laboratoire de M\'ecanique de Lille CNRS/UMR 8107, Universit\'e
  Lille 1 - Science et Technologie and Polytech'Lille, Cit\'e
  Scientifique Av.~P.~Langevin 59650 Villeneuve d'Ascq, France.}
\email[Corresponding author: ]{enrico.calzavarini@polytech-lille.fr}

\author{Federico Toschi}  
% \affiliation{International Collaboration for Turbulence Research}
\affiliation{International Collaboration for Turbulence Research,
  Dept.\ Physics and Dept.\ Mathematics \& Computer Science and
  J. M. Burgers Centre for Fluid Dynamics,\\Eindhoven University of
  Technology, 5600 MB Eindhoven, The Netherlands and\\ CNR-IAC, Via dei
  Taurini 19, 00185 Rome, Italy.}

% \collaboration{International Collaboration for Turbulence Research}

\begin{abstract}
  High Reynolds numbers Navier-Stokes equations are believed to break
  self-similarity concerning both spatial and temporal properties:
  correlation functions of different orders exhibit distinct
  decorrelation times and anomalous spatial scaling properties. Here,
  we present a systematic attempt to measure multi-time and
  multi-scale correlations functions, by using high Reynolds numbers
  numerical simulations of fully homogeneous and isotropic turbulent
  flow. The main idea is to set-up an ensemble of probing stations
  {\it riding the flow}, i.e. measuring correlations in a reference
  frame centered on the trajectory of distinct fluid particles (the
  quasi-Lagrangian reference frame introduced by Belinicher \& L'vov, Sov. Phys. JETP {\bf 66}, 303
(1987)). In this way we reduce the
  large-scale sweeping and measure the non-trivial temporal dynamics
  governing the turbulent energy transfer from large to small scales.
  We present evidences of the existence of {\it dynamic multiscaling
  }: multi-time correlation functions are characterized by an infinite
  set of characteristic times.
 
\end{abstract}

\date{\today}


\maketitle
\section{Introduction}
A comprehensive theory of the Eulerian and Lagrangian statistical
properties of turbulence is one of the outstanding open problems in
classical physics \cite{frisch:1995}.  Most studied quantities
concerns either measurements performed at the same time in multiple
positions (Eulerian measurements)
\cite{sreenivasan:1997,gotoh:2009}, or along one or several particles
moving with the flow (Lagrangian measurements)
\cite{yeung:2002,toschi:2009,benzi:jfm:2010}.  
The latter are optimal to study  temporal properties of the
underlying turbulent flows \cite{mordant:2001,arneodo:2008} but cannot
simultaneously disentangle also spatial fluctuations, being  based on 
single point, as for the case of acceleration
\cite{laporta:2001,sawford:2003, biferale:2005a,berg:2006a},
or on a evolving set of scales as for  two-particles
\cite{sawford:2001,biferale:2005b,bourgoin:2006,berg:2006b,salazar:2009}
and multi-particle dispersions \cite{biferale:2005c, xu:2008}.  On the
other hand, neither analytical control nor a firm phenomenological
description of fully developed turbulence can be obtained without a
solid understanding of the relation between spatio and temporal
fluctuations \cite{borgas:1993,lvov:1997,biferale:2004,chevillard:2006a,zybin:2008,zybin:2010,benzi:pre:2009,yu:2009,kleimann:2009,stresing:2010}.
In order to access unambiguously spatial and temporal fluctuations one
needs to set the reference scale and to get rid of the
large-scale sweeping in the same experimental --or numerical-- set-up
\cite{belinicher:1987,biferale:1999,mitra:2004,pandit:2008}. The idea
here is to exploit numerical simulations to define a set of probes
flowing with the wind, moving on a reference frame sticked with a
representative fluid particle.  In such a reference frame, we can
access velocity fluctuations over different spatial resolution,
together with their temporal evolution, without being affected by the
large-scale sweeping \cite{wan:2010}.  The interest of these
measurements is twofold. First, all attempts to break the theoretical
deadlock in turbulence have been hindered by the difficulties in
closing both spatial and temporal fluctuations
\cite{kraichnan:1974,lvov:1997,zybin:2010} (notice that the main
theoretical breakthrough in turbulent systems have been
obtained where temporal fluctuations are uncorrelated
\cite{falkovich:2001}). Second,  small-scale parametrizations used
for sub-grid turbulent closure call for more and more refined
phenomenological understanding of spatial and temporal fluctuations
\cite{meneveau:2000}.\\
In this article, we  show that quasi-Lagrangian measurement are able to remove the sweeping
effect, revealing that  correlation times in the Eulerian and Lagrangian
frame scale differently. Moreover, Lagrangian properties posses
a dynamical multi-scaling, i.e. different correlation functions
decorrelated with different characteristic time scales.  
 The locality in space and time of the energy cascade is supported
by studying the delayed peak in multi-time and multi-scale
correlations.
The main result we present is the confirmation of a theory
\cite{lvov:1997} where a bridge between spatial and temporal
intermittency was made by means of a refinement of the multifractal
phenomenology (originally proposed for the spatial statistics of
Eulerian velocity in \cite{parisifrisch:1985}). Energy is transferred
downscale with intermittent temporal fluctuations, and an associated
infinite hierarchy of decorrelation times. Temporal fluctuations
become wilder and wilder by decreasing the scale.

The article is organized as follows. 
In sec. II we present the numerical methods. In sec. III we show the result for single-scale multi-time correlation functions, for the bridge relation between temporal and spatial scaling exponents  and for multi-time and multi-scale correlation functions. 
%%%%%%%%%%
\begin{table*}
\begin{center}
\begin{tabular}{ccccccccccccccc}
  \hline 
  $N^3$ & $\delta x$ & $\delta t$ & $\nu$ & $\overline{\varepsilon}$ & $t_{tot}$& $N_p$ &  $M$ &
  $\eta$ & $\tau_{\eta}$ & $u_{rms}$ & $ \lambda $ &$L$& $T$  & $Re_{\lambda}$\\
  \hline
  $256^3$  & $2.4\!\cdot\!10^{-2}$   & $5\!\cdot\!10^{-4}$ & $3\!\cdot\!10^{-3}$ & $1.0$ & 110 & $3.2\!\cdot\!10^{4}$ & $20$ &
  $ 1.28\!\cdot\!10^{-2}$ & $5.48\!\cdot\!10^{-2}$ & $1.41$  & $3.00\!\cdot\!10^{-1}$& $4.24$& $3.00$ &  $141$\\
  \hline
  $512^3$  & $1.2\!\cdot\!10^{-2}$   & $4\!\cdot\!10^{-4}$ & $2.05\!\cdot\!10^{-3}$ & $0.9$ & 12 & $1.024\!\cdot\!10^{5}$ & $20$&
  $9.89\!\cdot\!10^{-3}$ & $4.77\!\cdot\!10^{-2}$ & $1.40$  & $2.58\!\cdot\!10^{-1}$& $4.56$& $3.26$ &  $176$\\                                  
  \hline
\end{tabular}
\caption{DNS parameters: $N$ is the number of grid points per spatial
  direction; $\delta x = 2\pi/N$ and $\delta t$ are the spatial and
  temporal discretization; $\nu$ is the value of kinematic viscosity;
  $\overline{\varepsilon}$ the mean value of the energy dissipation
  rate. $t_{tot}$ is the total simulation time; $N_p$
  total number of fluid tracers; $M$ number of  probes at fixed distances
  from  tracer particle;
  $\eta = (\nu^3/\overline{\varepsilon})^{1/4} $ and $\tau_{\eta} =
  (\nu/\overline{\varepsilon})^{1/2}$ the Kolmogorov dissipative
  spatial and temporal scales, $u_{rms}= \left( \overline{ \langle u_i
      u_i \rangle}_V/3 \right)^{1/2}$ the single-component
  root-mean-square velocity, $\lambda = (15\ \nu\ u_{rms}^2 /
  \varepsilon )^{1/2}$ the Taylor micro-scale, $T =
  (3/2)u_{rms}^2/\overline{\varepsilon}$ and $L= u_{rms}\, T$
  large-eddy-turnover temporal and spatial scales; $Re_{\lambda} =
  u_{rms}\ \lambda /\ \nu$ the Taylor scale based Reynolds
  number.\label{table1}}
   \vspace{-0.8cm}
\end{center}
\end{table*}
%%%%%%%%%%%%%%%%%%

\section{Methods}
\subsection{Sweeping and quasi-Lagrangian reference}
\begin{figure}[!t]
\vspace{-0.6cm}
  \begin{center}
    \includegraphics[width=1.0\columnwidth]{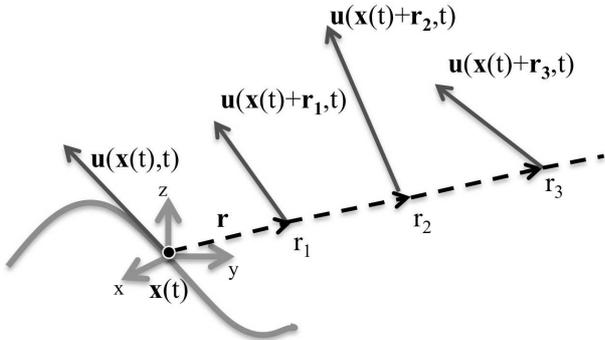}
    \caption{Sketch of the ${\bm r}_i$ probes ``riding the flow'',
      i.e., at fixed positions in a reference frame moving sticked
      with a representative fluid particle $\bfx(t)$.}
    \label{fig:draw}
        \vspace{-0.5cm}
  \end{center}
\end{figure}

The difficulty in studying the temporal correlations in turbulence is
associated with the sweeping of small scales by means of
larger ones. In the case of a flow with a ``large'' mean velocity
$\bfU$, fluctuations $\bfu' \equiv \bfu - \bfU$ are almost passively
transported in space via the Galilean transformation $\bfx'(t) = \bfx
+ \bfU t$.  This property, dubbed Taylor frozen flow hypothesis, is
commonly used in experiments to remap the hot-wire probe readings
(which are done in time) to measurements in space and is supposed to be
valid as long as turbulence levels are small ($|\bfu'|/|\bfU| \ll 1$). In
order to study the temporal evolution of velocity differences at a
given scale $r$ it is hence necessary to get rid of large-scale
(infra-red) effects. This can be done by \textit{``riding the flow''},
i.e. by sticking the origin of the reference frame on the position of
a fluid parcel moving in the flow (see sketch in fig.~\ref{fig:draw}).
Such reference frame, introduced by Belinicher and L'vov \cite{belinicher:1987}, is called
quasi-Lagrangian.  Because of this difficulty, measures of the
quasi-Lagrangian type have been performed only numerically at moderate
Reynolds and for small-scale quantities \cite{schumacher:2010} or at
high Reynolds in turbulence models, such as shell models (where sweeping is absent \cite{biferale:1999,lohse:2000,mitra:2004,pandit:2008}).
%%%%%%%%%%%%%%
%%%%%%%%%%%%%%%

\subsection{Numerical methods} 
In the present study we measure multi-scale and multi-time
correlations in a quasi-Lagrangian reference frame from fully resolved
high-statistics three-dimensional Direct Numerical Simulations (DNS)
of homogeneous and isotropic turbulence. We evolve the incompressible
Navier-Stokes (NS) equations:
\begin{equation} 
\partial_t \bfu + (\bfu \cdot
  \nabla) \bfu = - \nabla p +\nu \Delta \bfu + \bff\,, \quad \nabla
  \cdot \bfu = 0\,, \,\label{eq:ns}
\end{equation} 
in a cubic three-periodic domain via a pseudo-spectral algorithm and
$2^{nd}$ order Adams-Bashforth time marching.  The forcing $\bff$,
defined as in \cite{lamorgese:2004}, acts only on the first two shells
in Fourier space ($|\bfk| \leq 2$) and keeps constant in time the
total (volume averaged) injected power, $\langle \bff \cdot \bfu
\rangle_V= \textrm{const.}$ We report data coming from a set of
simulations with $N^3=256^3$ and $512^3$, corresponding to
$\mathrm{Re}_\lambda\, \simeq\,140$ and $180$ respectively (see Table
\ref{table1} for relevant parameters characterizing the flows).  The
simulation e.g. at $\mathrm{Re}_\lambda\, \simeq\,140$ has been
carried on for $40$ Eulerian turnover times,
$T=(3/2)u_{rms}^2/\overline{\varepsilon}$.  We also integrated
numerically $N_p=3.2 \cdot 10^4$ tracers evolving with the local
Eulerian velocity field $\dot{\bfx}(t) = {\bfu(\bfx(t),t)}$.  At fixed
temporal intervals % of $10 \delta t \simeq \tau_{\eta}/11$
we evaluate the fluid velocity also at ${\bm x}(t) + {\bm r}_i$, with
$i = 0,\ldots, M$ (spatial distances from each tracer). The vectors
are chosen to be always along one fixed direction, $\hat{\bfr}$, and
are logarithmically spaced in the range between zero and half of the
box-size (we use $M=20$), see figure \ref{fig:draw}.  Similar
measurements are done also at fixed positions uniformly spaced in the
fluid domain.  These two set of data are denoted respectively as
quasi-Lagrangian ($\cL$) and Eulerian ($\cE$).
%
%%%%%%%%%%%%%%
\begin{figure}[!tb]
  \begin{center}
    \includegraphics[width=\columnwidth]{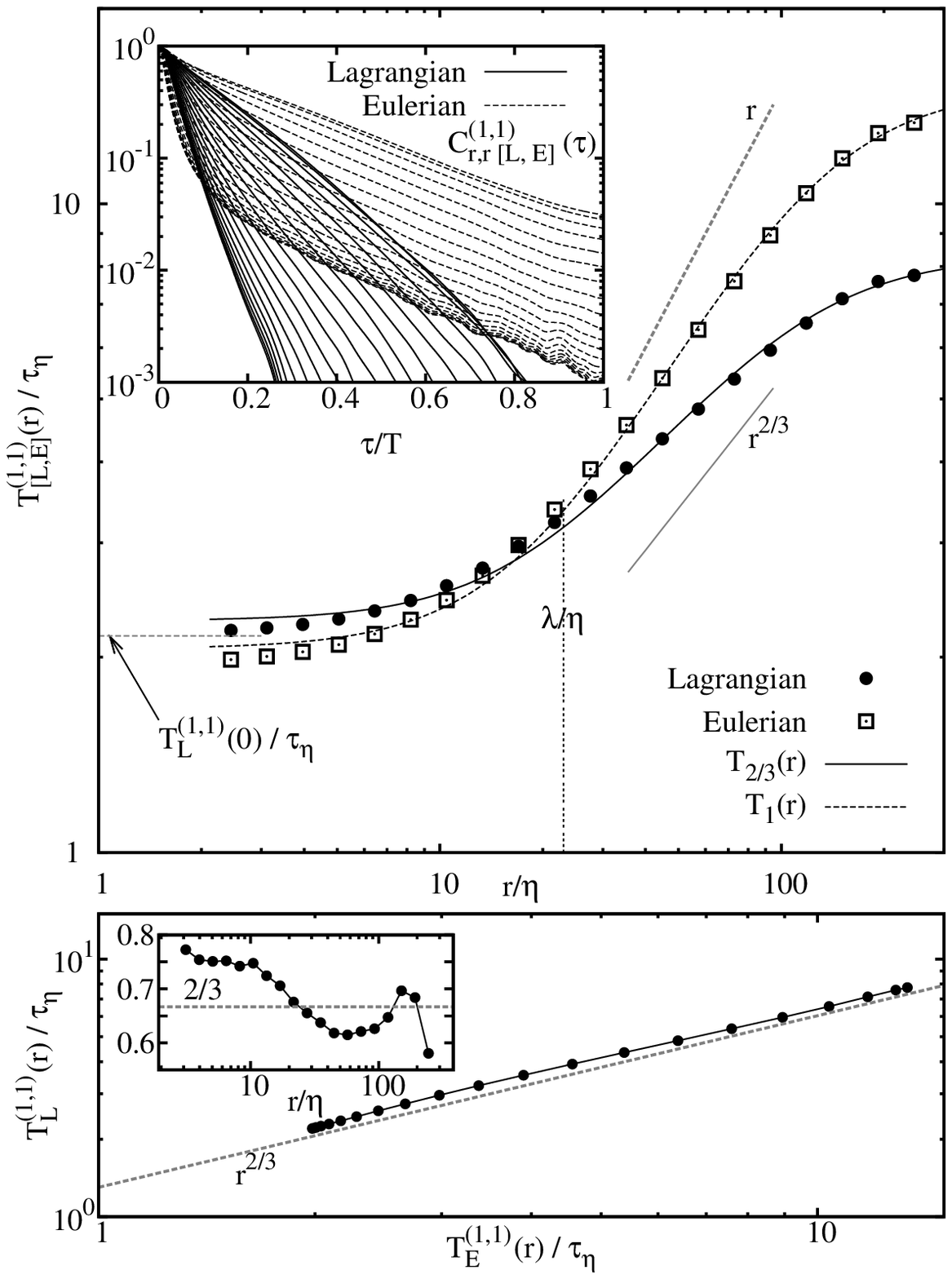}
    \caption{ Top-inset: Correlation functions of single-scale
      velocity differences $C_{r,r [\cL ,\cE]}^{(1,1)}(\tau)$ as a
      function of the time delay $\tau$, in the (quasi-)Lagrangian
      (solid line) and Eulerian (dashed line) frame of
      reference. Top-main panel: Single-scale integral correlation
      time $T^{(1,1)}_{\cL}(r)$ ($\bullet$) and $T^{(1,1)}_{\cE}(r)$
      ($\Box$) estimated from Lagrangian and Eulerian
      $C_{r,r}^{(1,1)}(\tau)$ correlations. The straight lines
      indicate slopes $2/3$ and $1$.  The fit using the function
      $T_{n}(r)$ discussed in the text with $n=2/3$ (solid) and $n=1$
      (dashed line) are also shown.  Vertical dotted line reports the
      value $\lambda \simeq 20 \eta$ for which both correlation times
      are $\sim \tau_{\eta}$. The horizontal line show the integral
      correlation time of longitudinal velocity gradients
      $T^{(1,1)}_{\cL}( \nabla_{||} \bfu ) = 2.15 \ \tau_{\eta}$ along
      Lagrangian trajectories, which corresponds to the
      small-scale(ultra-violet) limit, $\lim_{r \to 0}
      T^{(1,1)}_{\cL}(r)$. Bottom main panel: Extended Self Similarity plot:
      $T^{(1,1)}_{\cL}(r)\ vs.\ T^{(1,1)}_{\cE}(r)$
      and its local slope: $d \log{ T_{\cL}^{(1,1)}(r)}/ d
      \log{T_{\cE}^{(1,1)}(r)}$ (Bottom inset). The inertial range
      behavior $\sim r^{2/3}$ is reported in both panels.
    } \label{fig1} \vspace{-0.2cm}
  \end{center}
\end{figure}
%%%%%%%%%%%%%%%
\subsection{Notations and Measurements} 
We focus our attention on the longitudinal increments of velocity
differences across a displacement $\bfr$:
\begin{equation}
  \delta  u_{\bfr}(\tilde{\bfx} ,t) = \left( \bfu(\tilde{\bfx} + \bfr,t) - \bfu(\tilde{\bfx},t) \right)\cdot \hat{\bfr}.
\end{equation}
Notice that we have adopted a unifying notation, for us $\tilde{\bfx}$
can represent either a fixed point in space $\tilde{\bfx}=\bfx_0$ or a
point following a fluid particle: $\tilde{\bfx}=\bfx(t)=\int_{t_0}^{t}
\bfu(\bfx(t|{\bfx}_0,t_0),t)\, dt + \bfx_0$ (a trajectory passing from
$\bfx_0$ at time $t_0$).  We distinguish between the two cases by the
superscript labels: $\delta u_{\bfr}^{\cE}(\tilde{\bfx},t)$ or $\delta
u_{\bfr}^{\cL}(\tilde{\bfx},t)$.  Note that $\overline{ \left( \delta
    u_{\bfr}^{\cE}\right)^p}$, with overbar denoting time average, is
the usual Eulerian structure function of order $p$, furthermore by
means of ergodicity it can be proved that $\overline{ \left( \delta
    u_{\bfr}^{\cE}\right)^p} = \overline{ \left( \delta
    u_{\bfr}^{\cL}\right)^p}$ (therefore for such a quantity $\cE$,
$\cL$ labels will be dropped).
We define the generic multi-scale, multi-time correlation functions
\cite{lvov:1997}:
\begin{equation}
  \label{eqn_msmt}
  C_{R,r\, [\cL, \cE]}^{(q,p-q)}(\tau) = \frac{ \overline{ \left( \delta u_{\bfR}^{[\cL,\cE]}(t) \right) ^q \cdot \left( \delta u_{\bfr}^{[\cL,\cE]}(t+\tau)\right)^{p-q}}}{ \overline{( \delta u_{\bfR}(t) )^q \cdot ( \delta u_{\bfr}(t) )^{p-q} }  }, 
\end{equation}
where $\bfR$ and $\bfr$ denote separation vector fixed in space and
with different magnitude.  Note that the $\cL,\cE$ distinction must be
kept for the average of the multi-time product in the
numerator. Given the correlation functions we can define an Eulerian
and Lagrangian (integral) correlation time as follows
\cite{lvov:1997}:
\begin{equation}
  \label{eq:time} 
  T^{(q, p-q)}_{[\cL, \cE]}(R,r) = \int_{0}^{+\infty} C_{R,r\, [\cL, \cE]}^{(q,p-q)}(\tau)\ d\tau. 
\end{equation}
%%%%%%%%%%%%%%%%%%%%%%

\section{Results}
\subsection{Single-scale multi-time correlation}
We begin discussing the special case of a single-scale, multi-time
correlation, i.e.  $R = r$. 
%(hereafter $T^{(p)}_{[\cL, \cE]}(r) \equivT^{(q,p-q)}_{[\cL, \cE]}(r,r) $ for brevity).
% What is the expected
%behavior for $T^{(p)}_{\cE}(r)$ and $T^{(p)}_{\cL}(r)$ in a turbulent
%flow? 
Dimensional inertial-range scaling, $\overline{ \left( \delta
    u_{\bfr}\right)^p} \sim (\varepsilon\, r)^{p/3}$, provides the
following estimate for the turnover time of inertial eddies of size
$r$, $T^{(q,p-q)}_{\cL}(r) \sim r / \left( \overline{ \left( \delta
      u_{\bfr}\right)^p} \right)^{1/p} \sim r^{2/3}$.  On the contrary
the Eulerian correlation time -due to sweeping effect- can be
estimated by means of the typical velocity difference of the largest
eddy, which is proportional to the mean square root velocity, $\delta
u_{\bfL} \sim u_{rms}$. One has $T^{(q,p-q)}_{\cE}(r) \sim r / \left(
  \overline{ \left( \delta u_{\bfL}\right)^p} \right)^{1/p} \sim r /
u_{rms} \sim r$.  In the $r \to \eta$ limit both correlation times
tend to the dissipative scale $\tau_{\eta}$.
In Figure \ref{fig1} (inset) we show the behavior of
$C_{r,r}^{(1,1)}(\tau)$ for both Eulerian and quasi-Lagrangian
velocity differences and for separation scales $r \in[2.4,245]
\eta$. On the abscissa the time increment $\tau$ is made dimensionless
through the Eulerian large eddy turnover time $T$ (see
Tab.\ref{table1}). We clearly see that after a time $\sim T$ all the
correlations have decreased at least of a factor 50, supporting the
quality and convergence of our measurements. The main panel of Figure
\ref{fig1} shows  the integral correlation times both for the
Eulerian and quasi-Lagrangian case as computed from (\ref{eq:time}),
in a time integration window $[0,T]$.  The behavior is in qualitative
agreement with the expected scaling, the Lagrangian case being less
steep than the Eulerian one, however pure power-law scaling seems to
be hindered by finite Reynolds number and system finite-size effects.
To demonstrate this, we introduce a parametrization for the second
order spatial velocity structure functions, with dissipative and
large-scale cut-off (see also \cite{lohse:2000}): $T_{n}(r) = c_1
(1+(r/c_2)^2)^{n/2}(1+(r/c_3)^2)^{-n/2}$ with $n=2/3$ and $n=1$
respectively for the Lagrangian and Eulerian case. The parameters
$c_1,c_2,c_3$ represent the dissipative correlation time
scale, the dissipative  and large  cut-off
scales, respectively.  The good quality of the fit, shown in Fig. \ref{fig1},
supports our hypothesis. Plotting the Lagrangian correlation time as a
function of the Eulerian one, a procedure similar to the Extended Self
Similarity (ESS) \cite{benzi:1996}, does show a good scaling with
slope $0.64 \pm 0.02$ in the range $[20,200]\eta$, consistent with
$2/3$ (Fig. \ref{fig1}).  This finding again supports the idea that the
limited scaling in Figure \ref{fig1} is due to finite Reynolds
effects.

%\noindent 
\subsection{Intermittency and test of the bridge relations}
\begin{figure}[!t]
\vspace{-0.2cm}
  \begin{center}
    \includegraphics[width=1.\columnwidth]{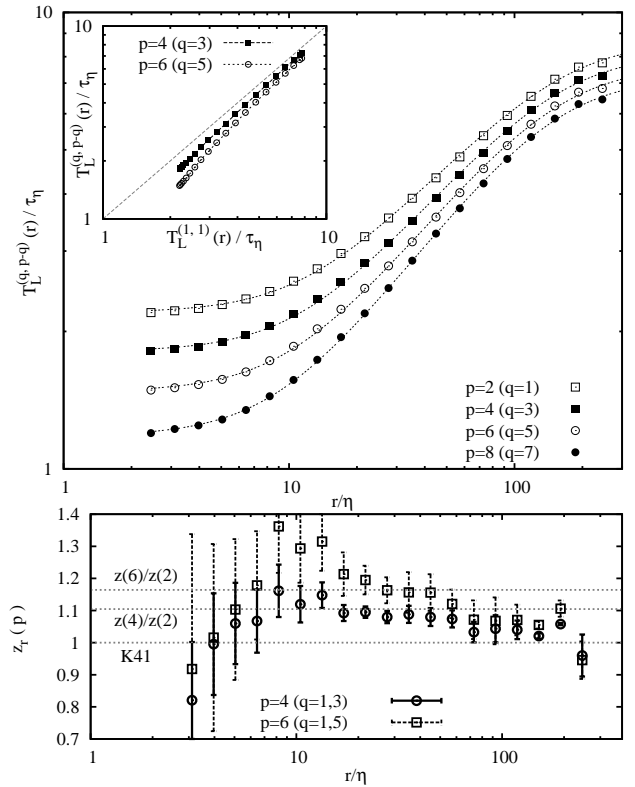}
    \caption{Test of dynamic multiscaling. Top panel: $T^{(q,p-q)}_{\cL}(r)$ $vs.$ $r/\eta$, with $p=2,4,6,8$ and $q=p-1$.
       Inset top panel: ESS-like plot of $T^{(q,p-q)}_{\cL}(r)\, vs.\, T^{(1,1)}_{\cL}(r)$ for $p=4,6$, $q=p-1$.  Bottom panel: local scaling exponents, 
$z_r(p)$, for $p=4,6$ and
      $q=p-1,1$. Central values are computed as the mean between  $q=p-1$ or $q=1$ for each $p$ value. Error bars gives the dispersion between the two
choices. 
Horizontal lines represents
      from bottom to top respectively the dimensional prediction
      $z(p)/z(2)=1$, $\forall p$ denoted as $K41$, and the BR values
      $z(4)/z(2) = 0.74/0.67$ and $z(6)/z(2) = 0.78/0.67$.
    }
    \label{fig2}
    \vspace{-0.2cm}
  \end{center}
\end{figure}

It is well known that Eulerian statistics shows intermittent
corrections to dimensional scaling. For example, for structure
function we have $\overline{ \left( \delta u_{\bfr}\right)^p} \sim r
^{\zeta(p)}$ where $\zeta(p)$ is a nonlinear convex function of $p$
\cite{frisch:1995}. In 1997, L'vov, Podivilov \& Procaccia
\cite{lvov:1997} provided a possible framework to encompass the
phenomenology associated to intermittency also to temporal
fluctuations. The idea consists in noticing that for time
correlations the structure of the advection term of the Navier-Stokes
equations suggests the relation: $T^{(q,p-q)}_{\cL}(r) \sim r / \left(
  \overline{\left( \delta u_{\bfr} \right)^{p}}\, / \,
  \overline{\left( \delta u_{\bfr}\right)^{p-1}} \right) \sim
r^{z(p)}$ \cite{lvov:1997}. Using the scaling for the Eulerian
quantities, $\overline{\left( \delta u_{\bfr} \right)^{p}} \sim
r^{\zeta(p)} $ one gets to the so-called \textit{bridge relations}
(BR) connecting spatial and temporal properties: $$z(p) = 1 - \zeta(p)
+ \zeta(p-1).$$ Similar idea have also been successfully applied to
connect the statistics of acceleration and velocity gradients
\cite{biferale:2004}.  Plugging the empirical values \cite{gotoh:2009,benzi:jfm:2010} for the Eulerian
exponents in the previous expression, one predicts $z(p) = 0.67, 0.74,
0.78, 0.80 (\pm 0.01)$ for the orders $p=2, 4, 6, 8$ respectively.  In
Figure \ref{fig2} (main top panel) the different moments of the
Lagrangian integral times, \textit{i.e.}  $T_{\cL}^{(q,p-q)}(r)$ (with
$p-q=1$), are shown versus the scale $r$. A steepening of the scaling
properties with increasing $p$ can be noticed.
%%%%%%%%%%%%%%
In order to enhance the quality of the measurements we resort to ESS
procedure by plotting a generic $T_{\cL}^{(q,p-q)}(r)$ versus
$T_{\cL}^{(1,1)}(r)$ (see inset of Figure \ref{fig2}) in log-scale.
It is now possible to define local scaling exponents as:
 $$ z_r(p) = d\log{T_{\cL}^{(q,p-q)}(r)}/ d \log{ T_{\cL}^{(1,1)}(r)},$$
 which according
to the BR  should scale as $z(p)/z(2)$ for $r$ in the
inertial range.  The result of this local scaling exponent analysis is
shown in Figure \ref{fig2} (bottom panel) for the orders $p=4,6$. 
Notice that the BR predicts the same scaling properties independently of $q$. In our numerics we find slightly different results for $q=1$ or $q=p-1$. The error
bars in bottom panel of fig. (\ref{fig2}) gives a quantitative estimate of the spread between the two results. 
 In the
inertial range $\sim[10,100]\eta$ we find  some deviation from the
K41 values $z(p)/z(2)=1$,  consistent with the BR predictions for $p=4,6$.
We notice that the predicted intermittent corrections are very small
and error bars large. Higher statistics and/or higher
Reynolds number may help in giving stronger confirmation to these
evidences.
%%%%%%%%%%%%%%
\begin{figure}[!t]
  \begin{center}
  \vspace{-0.2cm}
      \includegraphics[width=1.\columnwidth]{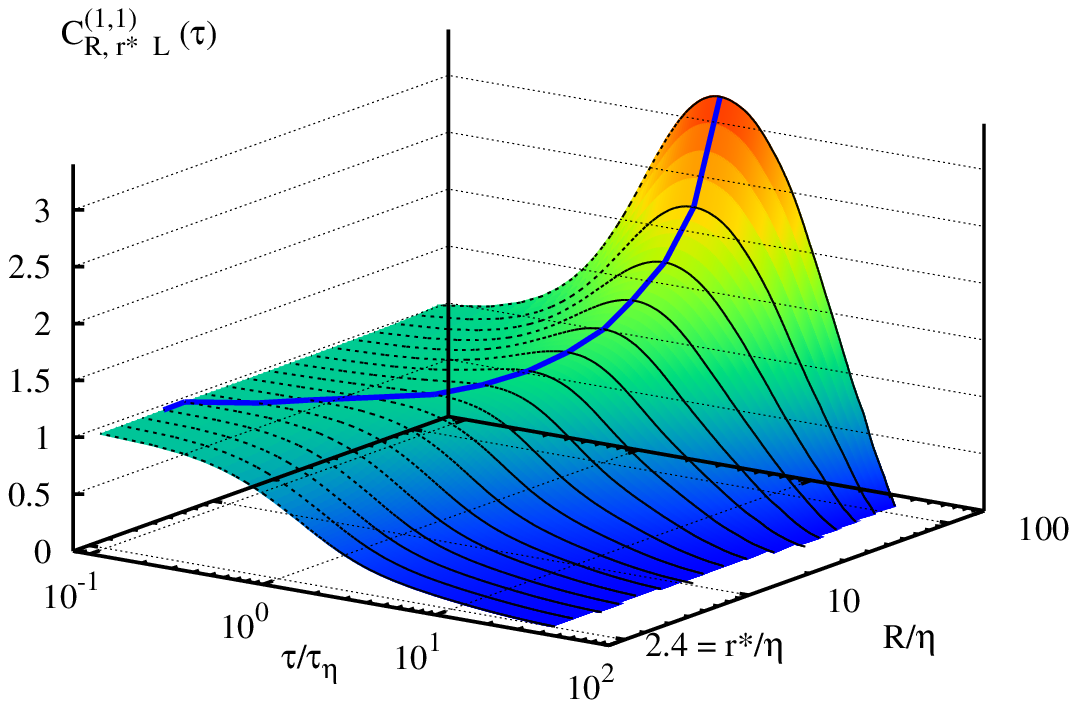}
    \includegraphics[width=1.\columnwidth]{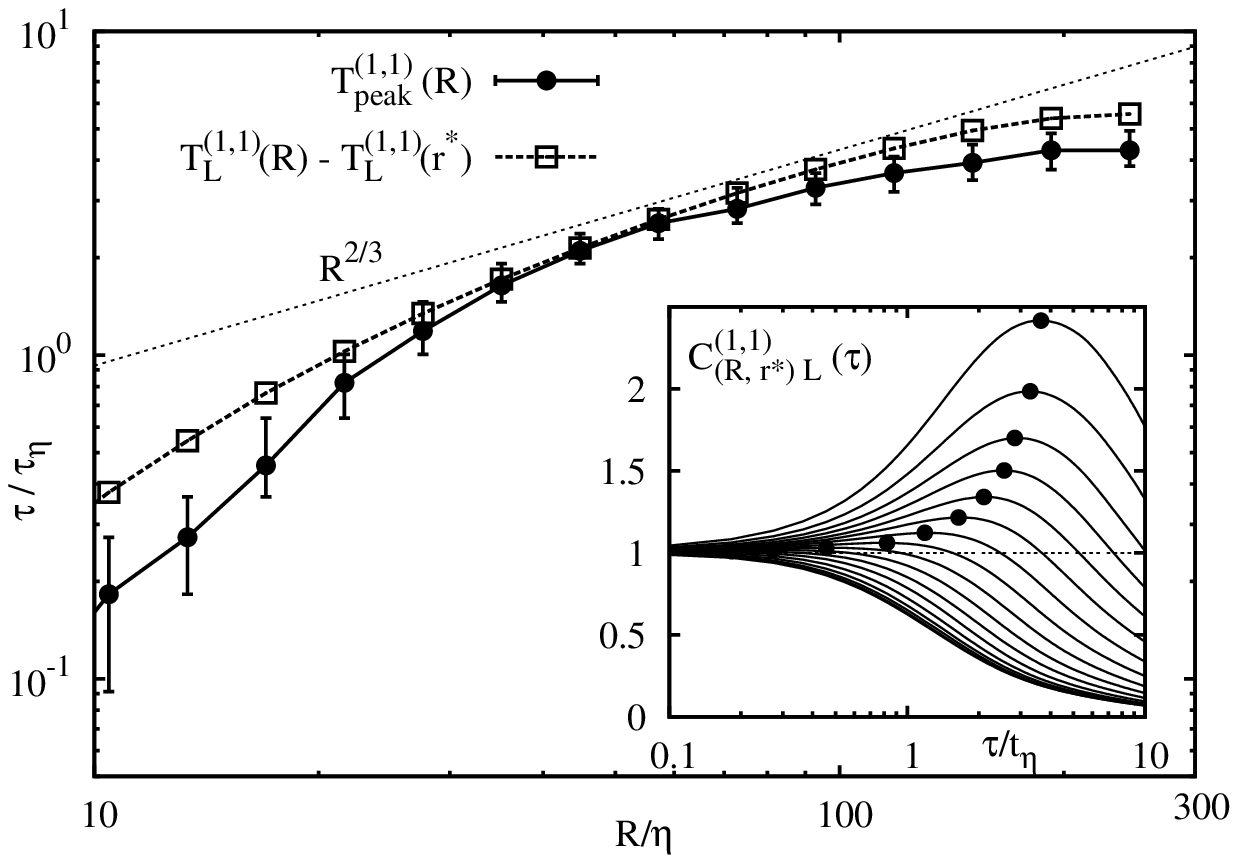}
    \caption{Top panel:
      $C_{R,r^{*}\, L}^{(1,1)}(\tau)$ for fixed  $r^{*}= 2.4 \eta$ at changing
$R \in [ r^{*}:120\eta]$. The solid line is a guide to the eyes connecting the peak for each given $R$. Inset bottom panel: Cut of   
$C_{R,r^{*}\, L}^{(1,1)}(\tau)$ at increasing $R$ (from bottom to top) as
a function of $\tau$.
      Symbol ($\bullet$) marks the position of the maximal correlation value
      (denoted as $T^{(1,1)}_{peak}(R)$) which increases for increasing
      values of $R$.  Bottom panel: Comparison between $T^{(1,1)}_{peak}(R)$
      and $T^{(1,1)}_{\cL}(R)- T^{(1,1)}_{\cL}(r^{*})$.  The inertial
      scaling $R^{2/3}$ is also drawn for comparison.
      }
    \label{fig3}
    \vspace{-0.8cm}
  \end{center}
\end{figure}
  %%%%%%%%%%%%%%% 
\subsection{Multi-scale multi-time correlation} 
We now focus on the most general case of multi-scale and multi-time
correlation functions in the Lagrangian frame. In particular, in the
correlation function (\ref{eqn_msmt}) we vary the large scale $R$
while the small scale $r$ is kept fixed $r \simeq \eta$. Note that the
velocity difference $\delta u_{\bfR}$ precedes in time the difference
$\delta u_{\bfr \simeq \eta}$.  We are therefore interested in the
time it takes for a velocity fluctuation to cascade down from a large
eddy (of size $R$) to the smallest one (of size $\sim \eta$).  In
Figure \ref{fig3} (top panel) we show the correlations
$C_{R,r^*\, \cL}^{(1,1)}(\tau)$ with $r^*=2.4\, \eta$ as defined from
Eqn.  (\ref{eqn_msmt}) (except for the fact that to enhance the
contrast instead of $\delta u_{\bfr}$ we used $|\delta u_{\bfr}|$ ).
The presence of a peak in $C_{R,r^*\, \cL}^{(1,1)}(\tau)$ for each given $R$,
defines a time, $T^{(1,1)}_{peak}(R)$, which increases for increasing values of
$R$.  The presence of the peak can be directly associated with the
time lapse it takes the energy to go down through scales from $R$ to
$r^*$, i.e. a direct evidence of temporal properties of the {\it  Richardson} turbulent cascade \cite{frisch:1995}. In the inset of the bottom panel we show the curves corresponding to  $C_{R,r^*\, \cL}^{(1,1)}(\tau)$, for each different $R$ at varying $\tau$.\\
The scaling behavior of the peak time $T^{(1,1)}_{peak}(R) \sim R^{2/3}$, shown in Fig. \ref{fig3} (bottom panel), 
is in agreement with what has been  measured for Fourier-space based quantities by Wan \textit{et al.} in \cite{wan:2010}.\\
Also in figure \ref{fig3} (bottom panel) a comparison of
$T^{(1,1)}_{peak}(R)$ with $T^{(1,1)}_{\cL}(R)- T^{(1,1)}_{\cL}(r^*)$ (as
computed for Fig. \ref{fig1}) is shown.  It is remarkable to note that
the amplitude and scaling of $T^{(1,1)}_{peak}(R)$ -coming from the
multi-scale correlation function- is close and compatible with
$T^{(1,1)}_{\cL}(R)-T^{(1,1)}_{\cL}(r^*)$ -coming from the single-scale
correlation functions. This finding provides a clean confirmation that energy is transferred down-scale in the quasi-Lagrangian reference frame.\\
  
\section{Conclusions}   
We presented an investigation of multi-scale and multi-time velocity correlations in
hydrodynamic turbulence in Eulerian and quasi-Lagrangian reference
frame.  
 Our main results are the following: i) We have demonstrated
that quasi-Lagrangian measurement are able to remove the sweeping
effect. The integral correlation times in the Eulerian and Lagrangian
frame are shown to scale differently. ii) Lagrangian properties posses
a dynamical multi-scaling, i.e. different correlation functions
decorrelated with different characteristic time scales.  iii) Bridge
relations connecting single-time multi-scale exponents with multi-time
single-scale exponents are valid, within numerical accuracy.
iv) The locality in space and time of the energy cascade is supported
by studying the delayed peak in multi-time and multi-scale
correlations.
Temporal fluctuations becomes larger and larger by going to smaller and smaller scales,  a phenomenon that
may even affect numerical stability criteria for time marching,
similarly to an effect concerning spatial resolution induced by
spatial intermittency \cite{schumacher:2007}. 
Some issues similar to the ones here discussed  have also been addressed in a recent numerical study
\cite{wan:2010} where evidences of the Lagrangian nature of the
turbulent energy cascade have been demonstrated by studying the
correlation  between energy dissipation and local energy
fluxes in the quasi-Lagrangian frame.  While the method followed in
\cite{wan:2010} requires a knowledge of the three-dimensional velocity
field, the approach proposed in the present manuscript needs only the
knowledge of the velocity at just a few points along a Lagrangian
trajectory: a measurements which may be already accessible in current
particle tracking experimental set-ups.

\textit{Acknowledgments} The authors acknowledge useful discussions with
R. Pandit and P. Perlekar. E.C. wish to thank J.-F. Pinton for
discussions and support to the initial stages of this work.  E.C. has
been supported also by the HPC-EUROPA2 project (project number:
228398) with the support of the European Commission - Capacities Area
- Research Infrastructures. We acknowledge COST Action MP0806 and computational support from
SARA (Amsterdam, The Netherlands), CINECA (Bologna, Italy) and CASPUR
(Rome, Italy).

%\bibliographystyle{prsty_withtitle}
%\bibliography{biblio_multiscaling}

\begin{thebibliography}{10}

\bibitem{frisch:1995}
U. Frisch, {\em Turbulence: The Legacy of A.~N.~Kolmogorov} (Cambridge
  University Press, Cambridge, 1995).

\bibitem{sreenivasan:1997}
K.~R. Sreenivasan and R.~A. Antonia, {\em The phenomenology of small-scale
  turbulence}, Ann. Rev. Fluid Mech. {\bf 29},  435  (1997).

\bibitem{gotoh:2009}
T. Ishihara, T. Gotoh, and Y. Kaneda, {\em Study of High-Reynolds Number
  Isotropic Turbulence by Direct Numerical Simulation}, Ann. Rev .Fluid Mech.
  {\bf 41},  165  (2009).

\bibitem{yeung:2002}
P.~K. Yeung, {\em Lagrangian investigations of turbulence}, Ann. Rev. Fluid
  Mech. {\bf 34},  115  (2002).

\bibitem{toschi:2009}
F. Toschi and E. Bodenschatz, {\em Lagrangian properties of particles in
  Turbulence}, Ann. Rev. Fluid Mech. {\bf 41},  375  (2009).

\bibitem{benzi:jfm:2010}
R. Benzi, L. Biferale, R. Fisher, D.~Q. Lamb, and F. Toschi, {\em Inertial
  range Eulerian and Lagrangian statistics from numerical simulations of
  isotropic turbulence}, J. Fluid Mech. {\bf 653},  221  (2010).

\bibitem{mordant:2001}
N. Mordant, P. Metz, O. Michel, and J.~F. Pinton, {\em Measurement of
  Lagrangian Velocity in Fully Developed Turbulence}, Phys. Rev. Lett. {\bf
  87},  214501  (2001).

\bibitem{arneodo:2008}
A. Arneodo, R. Benzi, J. Berg, L. Biferale, E. Bodenschatz, and \textit{et
  al.}, {\em Universal Intermittent Properties of Particle Trajectories in
  Highly Turbulent Flows}, Phys. Rev. Lett. {\bf 100},  254504  (2008).

\bibitem{laporta:2001}
A.~L. Porta, G.~A. Voth, A.~M. Crawford, J. Alexander, and E. Bodenschatz, {\em
  Fluid particle accelerations in fully developed turbulence}, Nature {\bf
  409},  1017  (2001).

\bibitem{sawford:2003}
B.~L. Sawford, P.~K. Yeung, M.~S. Borgas, P. Vedula, A.~L. Porta, A.~M.
  Crawford, and E. Bodenschatz, {\em Conditional and unconditional acceleration
  statistics in turbulence}, Phys. Fluids {\bf 15},  3478  (2003).

\bibitem{biferale:2005a}
L. Biferale, G. Boffetta, A. Celani, A. Lanotte, and F. Toschi, {\em Particle
  trapping in three-dimensional fully developed turbulence}, Phys. Fluids {\bf
  17},  021701  (2005).

\bibitem{berg:2006a}
J. Berg, {\em Lagrangian one-particle velocity statistics in a turbulent flow},
  Phys. Rev. E {\bf 74},  016304  (2006).

\bibitem{sawford:2001}
B. Sawford, {\em Turbulent relative dispersion}, Ann. Rev. Fluid Mech. {\bf
  33},  289  (2001).

\bibitem{biferale:2005b}
L. Biferale, G. Boffetta, A. Celani, B.~J. Devenish, A. Lanotte, and F. Toschi,
  {\em Lagrangian statistics of particle pairs in homogeneous isotropic
  turbulence}, Phys. Fluids {\bf 17},  115101  (2005).

\bibitem{bourgoin:2006}
M. Bourgoin, N.~T. Ouellette, H. Xu, J. Berg, and E. Bodenschatz, {\em The Role
  of Pair Dispersion in Turbulent Flow}, Science {\bf 311},  835  (2006).

\bibitem{berg:2006b}
J. Berg, B. Luthi, J. Mann, and S. Ott, {\em Backwards and forwards relative
  dispersion in turbulent flow: an experimental investigation}, Phys. Rev. E
  {\bf 74},  016304  (2006).

\bibitem{salazar:2009}
J.~P. L.~C. Salazar and L.~R. Collins, {\em Two-Particle Dispersion in
  Isotropic Turbulent Flows}, Ann. Rev. Fluid Mech. {\bf 41},  405  (2009).

\bibitem{biferale:2005c}
L. Biferale, G. Boffetta, A. Celani, B.~J. Devenish, A. Lanotte, and F. Toschi,
  {\em Multiparticle dispersion in fully developed turbulence}, Phys. Fluids
  {\bf 17},  111701  (2005).

\bibitem{xu:2008}
H. Xu, N.~T. Ouellette, and E. Bodenschatz, {\em Evolution of geometric
  structures in intense turbulence}, New J. Phys. {\bf 10},  013012  (2008).

\bibitem{borgas:1993}
M.~S. Borgas, {\em The Multifractal Lagrangian Nature of Turbulence}, Phil.
  Trans.: Physical Sciences and Engineering {\bf 342},  379  (1993).

\bibitem{lvov:1997}
V. L'vov, E. Podivilov, and I. Procaccia, {\em Temporal multiscaling in
  hydrodynamic turbulence}, Phys Rev E {\bf 55},  7030  (1997).

\bibitem{biferale:2004}
L. Biferale, G. Boffetta, A. Celani, B.~J. Devenish, A. Lanotte, and F. Toschi,
  {\em Multifractal statistics of Lagrangian velocity and acceleration in
  turbulence}, Phys. Rev. Lett. {\bf 93},  064502  (2004).

\bibitem{chevillard:2006a}
L. Chevillard and C. Meneveau, {\em Lagrangian Dynamics and Statistical
  Geometric Structure of Turbulence}, Phys. Rev. Lett. {\bf 97},  174501
  (2006).

\bibitem{zybin:2008}
K. Zybin, V. Sirota, A. Ilyin, and A. Gurevich, {\em Lagrangian statistical
  theory of fully developed hydrodynamical turbulence}, Phys. Rev. Lett. {\bf
  100},  174504  (2008).

\bibitem{zybin:2010}
K.~P. Zybin and V.~A. Sirota, {\em Lagrangian and Eulerian Velocity Structure
  Functions in Hydrodynamic Turbulence}, Phys. Rev. Lett. {\bf 104},  154501
  (2010).

\bibitem{benzi:pre:2009}
R. Benzi, L. Biferale, E. Calzavarini, D. Lohse, and F. Toschi, {\em Velocity
  gradients along particles trajectories in turbulent fows: the refined
  similarity hypothesis in the Lagrangian frame}, Phys. Rev. E {\bf 80},
  066318  (2009).

\bibitem{yu:2009}
H. Yu and C. Meneveau, {\em Lagrangian Refined Kolmogorov Similarity Hypothesis
  for Gradient Time-evolution in Turbulent Flows}, Phys. Rev. Lett. {\bf 104},
  084502  (2010).

\bibitem{kleimann:2009}
J. Kleimann, A. Kopp, H. Fichtner, and R. Grauer, {\em Statistics of a mixed
  Eulerian-Lagrangian velocity increment in fully developped turbulence},
  Physica Scripta {\bf 79},  055403  (2009).

\bibitem{stresing:2010}
R. Stresing and J. Peinke, {\em Towards a stochastic multi-point description of
  turbulence}, New Journal of Physics {\bf 12},  103046  (2010).

\bibitem{belinicher:1987}
V.~I. Belinicher and V.~S. L'vov, {\em A scale-invariant theory of developed
  hydrodynamic turbulence}, Sov. Phys. JETP {\bf 66},  303  (1987).

\bibitem{biferale:1999}
L. Biferale, G. Boffetta, A. Celani, and F. Toschi, {\em Multi-time,
  multi-scale correlation functions in turbulence and in turbulent models},
  Physica D {\bf 127},  187  (1999).

\bibitem{mitra:2004}
D. Mitra and R. Pandit, {\em Varieties of dynamic multiscaling in fluid
  turbulence}, Phys Rev Lett {\bf 93},  024501  (2004).

\bibitem{pandit:2008}
R. Pandit, S.~S. Ray, and D. Mitra, {\em Dynamic multiscaling in turbulence},
  Eur. Phys. J. B {\bf 64},  463  (2008).

\bibitem{wan:2010}
M. Wan, Z. Xiao, C. Meneveau, G.~L. Eyink, and S. Chen, {\em Dissipation-energy
  flux correlations as evidence for the Lagrangian energy cascade in
  turbulence}, Phys Fluids {\bf 22},  061702  (2010).

\bibitem{kraichnan:1974}
R.~H. Kraichnan, {\em On Kolmogorov's inertial-range theories}, J. Fluid Mech.
  {\bf 62},  305   (1974).

\bibitem{falkovich:2001}
G.~K. Falkovich~G and V. M, {\em Particles and fields in fluid turbulence},
  Rev. Mod. Phys. {\bf 73},  913  (2001).

\bibitem{meneveau:2000}
C. Meneveau and J. Katz, {\em Scale-invariance and turbulence models for
  large-eddy simulation}, Ann. Rev. Fluid Mech. {\bf 32},  1  (2000).

\bibitem{parisifrisch:1985}
G. Parisi and U. Frisch, {\em On the singularity structure of fully developed
  turbulence}, Turbulence and Predictability in Geophysical Fluid Dynamics,
  edited by M. Ghil, R. Benzi, and G. Parisi  84  (1985).

\bibitem{schumacher:2007}
J. Schumacher, {\em Sub-Kolmogorov-scale fluctuations in fluid turbulence},
  EuroPhys. Lett {\bf 80},  54001  (2007).

\bibitem{schumacher:2010}
J. Schumacher, B. Eckhardt, and C.~R. Doering, {\em Extreme vorticity growth in
  NavierÐStokes turbulence}, Physics Letters A {\bf 374},  861  (2010).

\bibitem{lohse:2000}
S. Grossmann, D. Lohse, and A. Reeh, {\em Multiscale correlations and
  conditional averages in numerical turbulence}, Phys. Rev. E {\bf 61},  5195
  (2000).

\bibitem{lamorgese:2004}
A.~G. Lamorgese, D.~A. Caughey, and S.~B. Pope, {\em Direct numerical
  simulation of homogeneous turbulence with hyperviscosity}, Phys. Fluids {\bf
  17},  015106  (2004).

\bibitem{benzi:1996}
R. Benzi, L. Biferale, S. Ciliberto, M. Struglia, and R. Tripiccione, {\em
  Generalized scaling in fully developed turbulence}, Physica D {\bf 96},  162
  (1996).

\end{thebibliography}

\end{document}